\documentclass[chicago,numberedappendix,tighten]{emulateapj}

\usepackage{amsmath,amssymb,natbib,graphicx}
\usepackage{color}
\usepackage{ulem}

\slugcomment{DRAFT \today}

\shorttitle{Radio Detection of Green Peas}
\shortauthors{Chakraborti et al.}

\begin{document}


\title{Radio Detection of Green Peas:\\
Implications for Magnetic Fields in Young Galaxies}


\author{Sayan Chakraborti, Naveen Yadav}
\affil{Department of Astronomy and Astrophysics, Tata Institute of Fundamental Research,\\
    1 Homi Bhabha Road, Colaba, Mumbai 400 005, India}

\author{Carolin Cardamone}
\affil{The Harriet W. Sheridan Center for Teaching and Learning, Brown University,\\
    Box 1912, 96 Waterman St, Providence RI 02912, USA}

\and

\author{Alak Ray}
\affil{Department of Astronomy and Astrophysics, Tata Institute of Fundamental Research,\\
    1 Homi Bhabha Road, Colaba, Mumbai 400 005, India}

\email{sayan@tifr.res.in}




\begin{abstract}
Green Peas are a new class of young, emission line
galaxies
that were discovered by citizen volunteers in the Galaxy Zoo
project.
Their low stellar mass, low metallicity and very high star formation
rates make Green Peas the nearby ($z\sim0.2$) analogs
of the Lyman-break Galaxies (LBGs)
which account for the bulk of the
star formation in the early universe ($z\sim2-5$). They thus
provide accessible laboratories in the nearby Universe for
understanding star formation, supernova feedback, particle
acceleration and magnetic field amplification in early galaxies.
We report the first direct radio detection of Green Peas with low
frequency GMRT observations and
our stacking detection with archival VLA
FIRST
data. We show that the radio emission implies that these
extremely young galaxies already have magnetic fields ($\gtrsim30\mu G$)
even larger than that of the Milky Way.  This is at
odds with the present understanding of magnetic field growth
based on amplification of seed fields by
dynamo action
over a galaxy's lifetime. Our observations strongly favor
models with pregalactic magnetic fields
at $\mu G$ levels.
\end{abstract}





\section{Introduction}
\begin{figure}[t]
\begin{center}
\includegraphics[width=0.7\columnwidth]{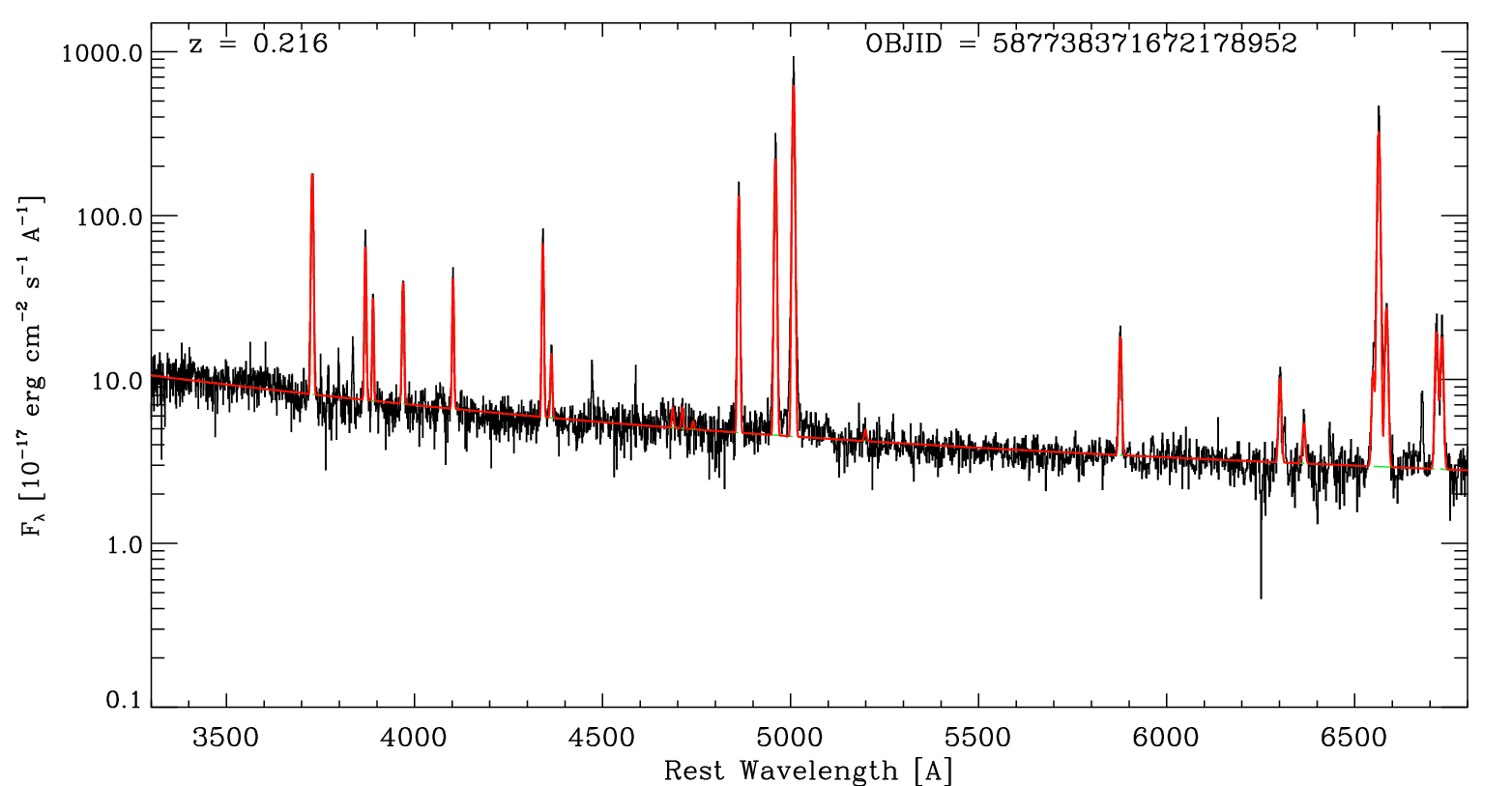}
\includegraphics[width=0.7\columnwidth]{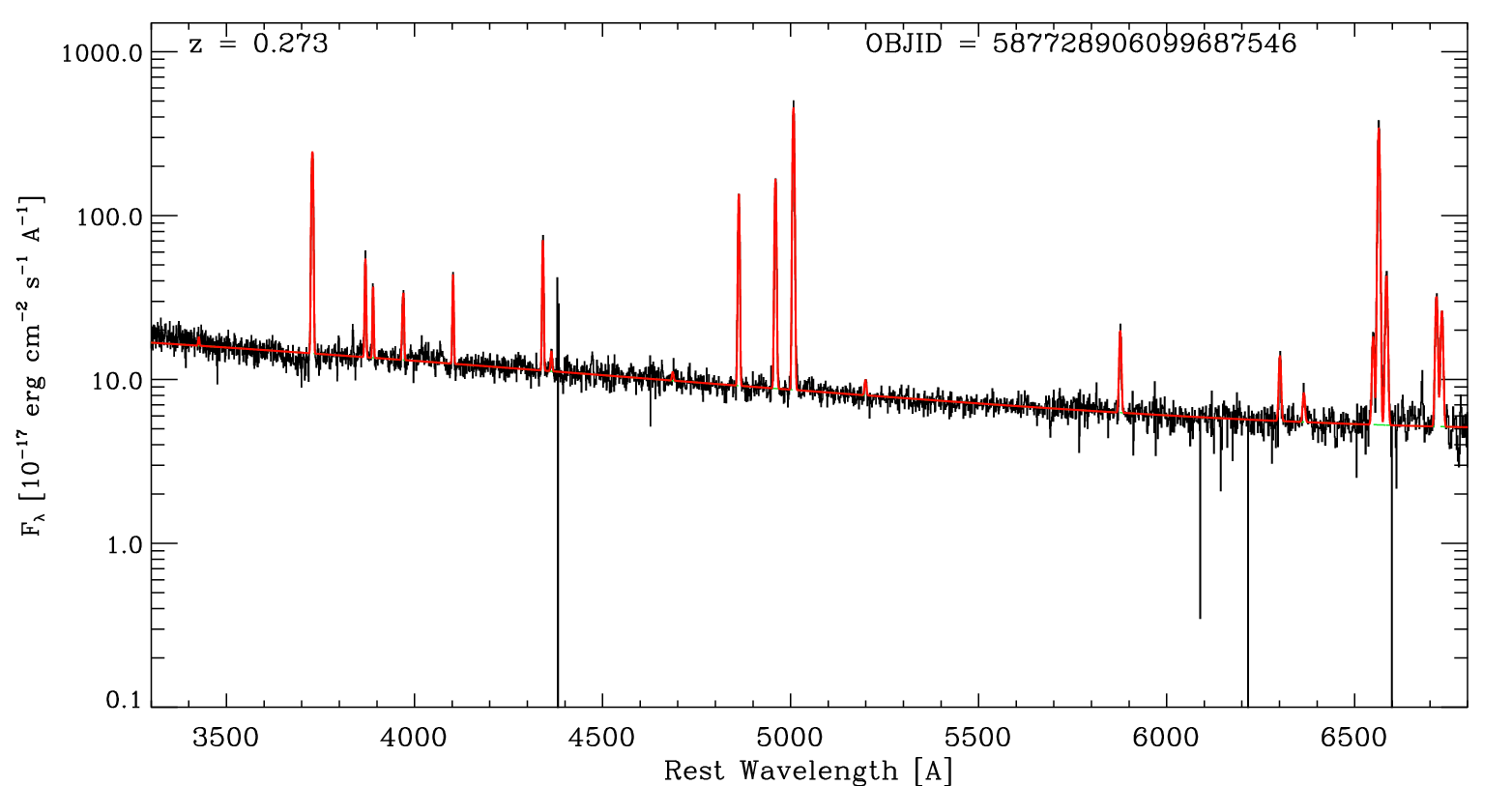}
\includegraphics[width=0.7\columnwidth]{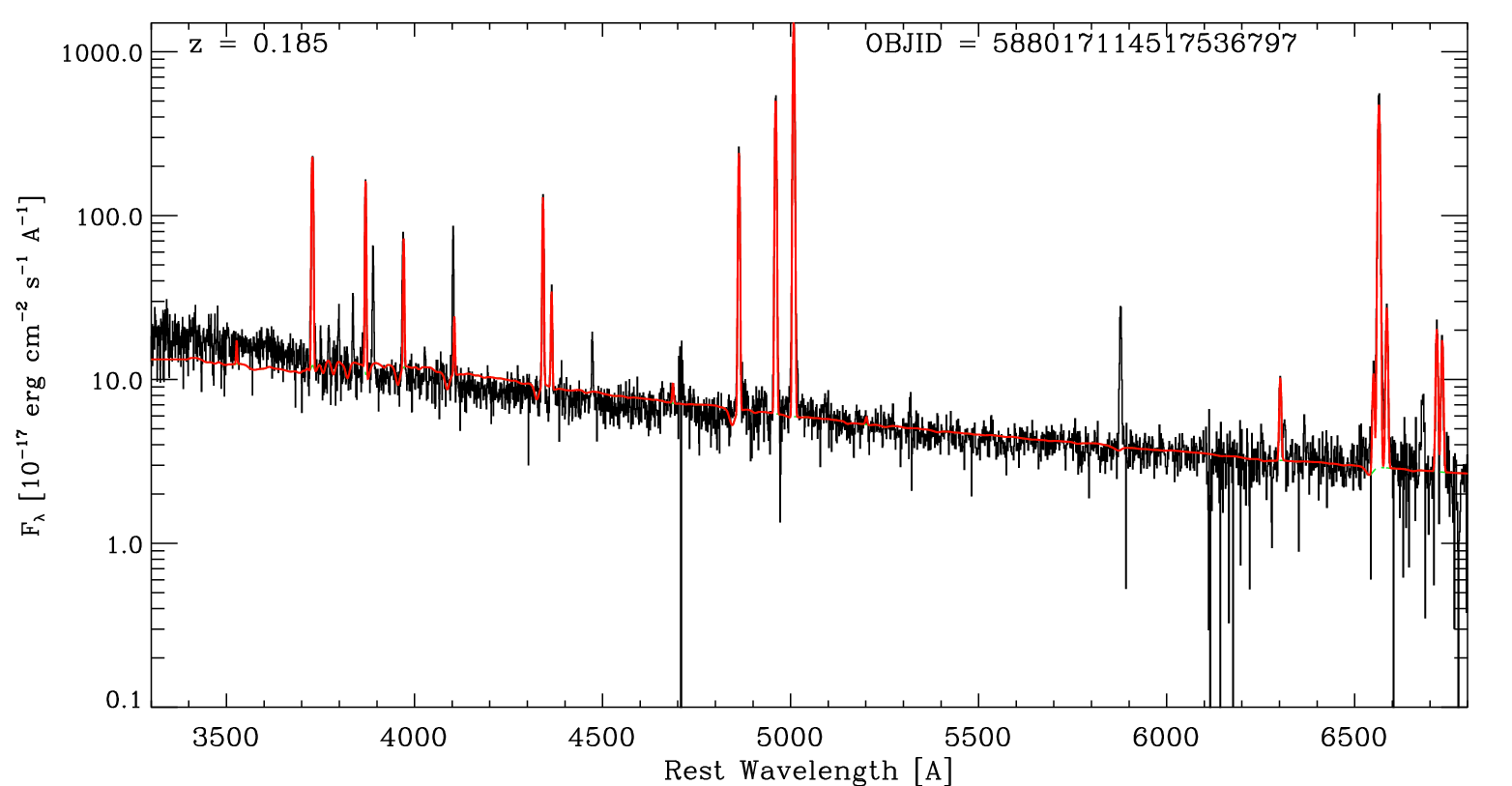}
\end{center}
\caption{SDSS spectra of three Peas, targeted with GMRT.
All spectra (in black) are de-redshifted and fitted (in red) with the Gas And
Absorption Line Fitting \citep[GANDALF;][]{2006MNRAS.366.1151S} that
simultaneously fits both stellar spectra and emission lines to
determine the metallicity, stellar mass and star formation rates.
See \citet{2009MNRAS.399.1191C} for details of the procedure.
\label{SDSS}}
\end{figure}

A new class of young emission line galaxies, named
Green Peas \citep{2009MNRAS.399.1191C}, was recently discovered
by volunteers in the Galaxy Zoo project \citep{2008MNRAS.389.1179L}.
They are particularly interesting because of their
small size and their extraordinarily
large [O{\sc iii}] equivalent width (up to $\sim$1000 \AA; see Fig. \ref{SDSS}),
which indicates a large population of hot young stars.
They turn out to be low mass galaxies (M$\sim10^{8.5}-10^{10} $ ${\rm M_{\odot}}$)
with high star formation rates ($\sim 10$ ${\rm M_{\odot}yr^{-1}}$).
Hence they have some of the the highest specific star formation
rates\footnote{$\dot M/M$ of up to $\sim10^{-8}$ ${\rm yr^{-1}}$ were
obtained by \citet{2009MNRAS.399.1191C}.
{Typical Peas have
$r\sim 20$, $u-r\sim 0.9$ and $z\sim 0.2$
\citet{2009MNRAS.399.1191C}.}}
seen in the local Universe. This indicates
very short stellar mass doubling times ($t_{\rm double}\sim10^8$yrs,
which is $\sim10^{-2}$ times the Hubble time), whereas
typical star forming galaxies have stellar mass doubling
times comparable to the Hubble time.
Due to their small reddening (${\rm E(B-V) \leq 0.25}$) and
luminous UV emission this class of star forming galaxies is similar
to UV luminous galaxies  observed at high redshift ($z\sim2-5$).
Their relatively low metallicities (log[O/H] + 12 $\sim$8.7) and
locations in under-dense environments suggest that Peas
started building their stellar content at $z\approx0.2$ by processes
similar to those experienced by high redshift Lyman-$\alpha$ emitters
and {
Lyman Break Galaxies \citep[LBGs; see][for a review]{2002ARA&A..40..579G}.}
It has been suggested by \citet{2010ApJ...715L.128A} that Peas
may have systematically larger [N/O] ratios, and therefore potentially
even lower metallicity than reported by \citet{2009MNRAS.399.1191C}. 

Though unresolved by the SDSS, {requiring typical
angular sizes $\lesssim$1'' which gives physical sizes
$\lesssim$5 kpc}, some Peas had serendipitous
Hubble Space Telescope (HST) observations.
Their patchy irregular appearance in resolved HST snapshots possibly
due to compact star forming regions \citep{2009MNRAS.399.1191C}, mimics
the morphologies of high redshift galaxies.
Given high star formation
rates, Peas are expected to host a large number of supernovae.
Supernovae accelerate
electrons to high energies, which may then emit synchrotron
radiation in radio bands. We have performed a stacking experiment
with archival Very Large Array (VLA) FIRST \citep{1995ApJ...450..559B}
data at 1.4 GHz to demonstrate that their fluxes are systematically different
from those of usual starburst galaxies but are statistically akin to those
of LBGs, which have systematically lower fluxes than local starbursts at a
given star formation rate \citep{2008ApJ...689..883C}.
We have followed up three Peas with deep observations
at 0.6 GHz using the Giant Metrewave Radio Telescope (GMRT).
In this letter we establish and discuss the properties of Peas as a
new class of sub-milliJansky sources in the GHz radio sky. 
{We use theoretical considerations to show that these
radio detections imply large magnetic fields for Peas.}
This raises puzzling questions about the origin and evolution
of magnetic fields in young galaxies.



\section{Radio Emission in Galaxies}
Radio emission from most normal\footnote{Here
normal \citep{1992ARA&A..30..575C} refers to nearby galaxies in which
the radio emission is not dominated by a central super-massive blackhole.
Seyfert Peas \citep{2009MNRAS.399.1191C} are excluded from this study and
we only concentrate on Star Forming Peas.}
galaxies is due to synchrotron radiation of relativistic electrons and
free-free emission of ionized hydrogen (H{\sc ii}) regions
\citep{1992ARA&A..30..575C}.
Both of are produced by massive stars ($M \gtrsim 8 M_\odot$)
which end their bright but short lives ($\lesssim 3 \times 10^7$ yr) in
supernovae. During their lifetimes these stars ionize the surrounding 
Interstellar Medium (ISM) and create H{\sc ii} regions.
Shocks produced by supernovae accelerate a majority of the
relativistic electrons in normal galaxies.
These electrons have short lifetimes ($\lesssim 10^8$ yr);
hence radio emission in a normal galaxy traces recent
star formation activity.

The non thermal
{radio brightness of}
a typical star forming galaxy
is given by \citet{2002ApJ...568...88Y} as
\begin{equation}
S_{\rm nth}(\nu) = 25~f_{\rm nth} \nu^{-\alpha}
\left({{\rm SFR}\over{\rm M_\odot yr^{-1}}}\right) {D_L}^{-2}~~{\rm Jy},
\label{eq:nth2}
\end{equation}
where $D_L$ is luminosity distance in Mpc. Variations in
the normalization of the nonthermal synchrotron emission are
traced by $f_{\rm nth}$, which is $\sim1$ for the Milky Way and local
starbursts \citep{2002ApJ...568...88Y}.
{With high star formation rates, young star forming
galaxies like LBGs and Peas are expected to have some nonthermal
radio emission. However, to date no direct radio emission has been reported.
In LBGs, stacking techniques \citep{2008ApJ...689..883C} have been used to
overcome large distances. With Peas, we have a
local sample to search for their radio emission both by
stacking existing VLA data (Section \ref{VLA}) and by additional
observations with GMRT (Section \ref{GMRTOBS}).}

\begin{figure}[t]
\begin{center}
\includegraphics[width=0.8\columnwidth]{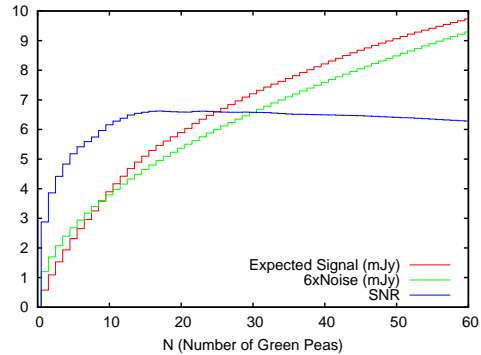}
\end{center}
\caption{Expected Signal (in mJy), $6\sigma$ detection threshold (in mJy)
and Signal to Noise Ratio (SNR) for a simulated stacking experiment with
N brightest Peas. Note that the SNR rises and then falls
with N, so stacking all available Peas is not optimal.
We stack $2^5$ Peas in this work.
\label{SNR}}
\end{figure}

\section{VLA FIRST Stacking Detection}
\label{VLA}
Highest estimated fluxes for Peas fall
just short of the VLA FIRST
survey threshold {of $\sim1$ mJy \citep{1995ApJ...450..559B}}
at 1.4 GHz, so they
need to be detected by stacking. {See
\citet{2007ApJ...654...99W} for extensive discussion of techniques for
stacking FIRST images to detect sources below the survey threshold.
Here we use a summation based approach, rather than a median based
approach, as this gives higher Signal to Noise Ratio (SNR). This is
because, the convergence of the sample mean to the
population mean is unbiased and faster than the convergence of the sample median
to the population median \citep{kenney1957}.}
Conventional wisdom suggests that summing up N images of faint objects
produces a increase in SNR by $\sqrt{N}$
This implies that stacking all available objects is the best strategy. However,
this strategy can be bettered if some flux estimate is available. We estimate
the expected fluxes of all
{80 spectroscopically identified
star forming}
Peas {from} \citet{2009MNRAS.399.1191C} using Eq \ref{eq:nth2} with
$f_{\rm nth}=1$ in a flat $\Lambda$CDM cosmology with $h=0.71$ and
$\Omega_M=0.3$, with star formation rates derived from their optical spectra
(See Figure \ref{SDSS}).
In this situation it is best to stack the N galaxies with the highest expected
fluxes. In Figure \ref{SNR} we show that the SNR rises rapidly in the range
$N\sim1-15$ and falls off slowly beyond $N\gtrsim30$. We therefore
stack $N=2^5$ (the nearest power of 2) of the brightest expected Peas.

\begin{figure}
\begin{center}
\includegraphics[width=0.8\columnwidth]{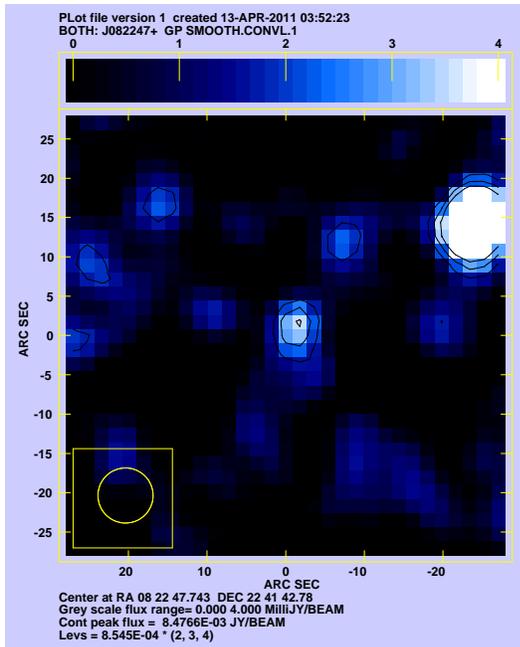}
\end{center}
\label{FIRST}
\caption{VLA FIRST detection of stacked flux from 32 Peas.
The displayed map is smoothed to a restoring beam size of $7^{\prime\prime}$.
The unresolved $4\sigma$ source in the centre is due to the Peas. The
only other significant detection, the resolved source to the right of
the image, is due to the unrelated {\it FIRST J133926.7+151655}.}
\end{figure}

The images of these $N$  regions
were cut out and summed in pairs, with this process being repeatedly
applied to the resulting $N/2$ maps until the final map
(Fig. \ref{FIRST}) was obtained.
This was done to avoid a sequential summation, which results in the
addition of large floating point numbers with small ones in the latter
steps, leading to accumulation of rounding errors. In our method, the
numbers being added are always comparable. The final image
(Fig. \ref{FIRST}) has noise
which is well described by a Gaussian with $\sigma=0.861$ mJy. The
stacked total flux is found to be $3.975$ mJy
instead of the expected $7.432$ mJy for $f_{\rm nth}=1$, {
summed over all N Peas using Eq \ref{eq:nth2} for a standard starburst
template}. The probability of a
statistical fluctuation of this size and larger is
$\sim9.2\times10^{-4}$.
This indicates that the average radio fluxes of Peas are
systematically suppressed, compared to usual starburst galaxies, by a
factor of $f_{\rm nth}\simeq0.53$.
Thus, as in LBGs \citep{2008ApJ...689..883C}, Peas
have a comparable but systematically lower flux when compared to local
starbursts.

{
\section{GMRT Observations and Results}
\label{GMRTOBS}
}
The result from stacking
demands confirmation at other frequencies,
especially in terms of direct radio detection of Peas. Hence,
we targeted the 3 most promising candidates, with
expected fluxes at the mJy level, using deep GMRT observations. We detect
two of them {as unresolved point sources} and put an upper
limit on the third. These results show that
not all Peas may be characterized by a single $f_{\rm nth}$. 
Observations {with $\sim3$ hours of on-source time
for each Pea},
were carried out at 617 MHz
(bandwidth 32MHz, split into 512 channels).
Visibility data {have} been analyzed using Astronomical Image
Processing System (AIPS) {\citep{1990apaa.conf..125G}}.

Initial {calibrations}
of the data {were} done using various flux calibrators and phase calibrators
(See Table \ref{GMRT}). The flux calibrators were
observed at the beginning and the end of the respective
observations in order to fix the antennae gains. The phase calibrators,
chosen within $15^\circ$ of the source position on the sky,
were observed every half an
hour to calibrate the phase drift due to ionospheric effects.
The data {were} edited for the removal of bad antennae, baselines
and any radio frequency interference (RFI).
The data {were} binned into seven channels to maximize SNR
without compromising on chromatic aberration. Target visibilities
were extracted from gain and bandpass calibrated data. The task
IMAGR was used on this data to obtain a preliminary dirty map. Cleaning
was done to remove the beam pattern for all $3\sigma$ and brighter sources.
Repeated self calibration and cleaning was done till it yielded
consistent solutions. The source flux was then extracted with the task
JMFIT from the final map, {with a typical resolution of
$\sim6$''}.
Table \ref{GMRT} lists sources, calibrators, fluxes and implied
magnetic fields.

\begin{deluxetable*}{lcllccc}
\tablecaption{\label{GMRT} GMRT Observations of Peas}
\tablewidth{0pt}
\tablehead{
\colhead{Pea} & \colhead{z} & \colhead{Flux Cal} & \colhead{Phase Cal} & \colhead{$S_\nu$} & \colhead{$B_{\rm D}$} & \colhead{$B_{\rm eq}$}\\
\colhead{(SDSS Object)} & \colhead{} & \colhead{(3C Source)} & \colhead{(J2000)} & \colhead{(mJy)} & \colhead{${\rm\;\mu G\;}$} & \colhead{${\rm\;\mu G\;}$}
}
\startdata
J082247.66+224144.1& 0.216 & 3C147, 3C286 	 & J0741+312	& $1.20\pm0.31$		& $81\pm14$		& $50\pm4$	\\
J074936.76+333716.3& 0.273 & 3C147		 & J0842+185	& $1.11\pm0.11$		& $82\pm5$		& $54\pm1$	\\
J142405.73+421646.3& 0.185 & 3C286		 & J1416+347	& $<0.45$		& $<49$			& $<35$	
\enddata
\tablecomments{Details of three {Peas observed at 617 MHz}:
$B_{\rm D}$ and $B_{\rm eq}$
are the magnetic fields derived using diffusion and equipartition
arguments. Uncertainties
are 1$\sigma$ statistical standard errors.
The {entries} for the last object represent the 3$\sigma$ upper
confidence limits.}
\end{deluxetable*}

\section{Electron Diffusion}
Relativistic electrons produce bulk of the GHz radio
emission in star forming galaxies. They must remain confined long enough
in the galaxy's magnetic field to significantly contribute to the
radio emission. { We use the criteria derived here, to constrain
the characteristic magnetic fields in Section \ref{magdiff}.} These electrons
traveling at speeds close to that of light get scattered by the
galaxy's magnetic field. Their diffusion coefficient is determined
by the structure of the magnetic field. We assume following
\citet{1964ocr..book.....G} that the
field structure is comprised of domains of radius $l_0$
(typically 0.3 kpc according to \citet{2010PhRvL.105i1101C})
with randomly oriented magnetic field $B$ in each of these domains.
For the range of feasible electron energies, the Larmor radius is
smaller than the magnetic coherence length, thereby providing an energy
independent diffusion coefficient { $D=l_0 c / 3$}.
Therefore the diffusion
timescale for relativistic electrons to escape from a typical Pea
of size 3 kpc \citep{2009MNRAS.399.1191C}, is {
\begin{equation}
 t_{\rm D} \simeq \frac{R^2}{D}\approx2.94\times10^5 {\rm\;yr\;}
\left(\frac{R}{3 {\rm kpc}}\right)^2
\left(\frac{l_0}{0.3 {\rm kpc}}\right)^{-1} .
\label{dif}
\end{equation}
}
This can be compared to the timescales of radiative processes
to gauge their relative importance.

\section{Inverse Compton scattering}
{Electrons may lose their energy before diffusing out of a
galaxy via Inverse Compton (IC) scattering.}
An explanation of the systematically lower fluxes of the
LBGs suggested by \citet{2008ApJ...689..883C}, is {IC}
cooling of relativistic electrons scattering off
Cosmic Microwave Background (CMB) photons.
If this is indeed the cause, the effect should be negligible for Peas
as the CMB radiation density in the Universe has fallen off considerably in the
epoch between LBGs ($z\sim2-5$) and Peas ($z\sim0.2$).
Electrons can lose their energy through IC.
IC has the same dependence on the photon energy density
as synchrotron has on the magnetic field energy
density \citep{1979rpa..book.....R}. This implies the IC energy
loss timescale;
\begin{align}
 t_{\rm IC} = & \frac{E}{|dE/dt|_{\rm IC}} \\
\simeq & 1.29\times10^8 {\rm\;yr\;}
\left(\frac{U_{\rm rad}}{10^{-12} {\rm erg cm^{-3}}}\right)^{-3/4}
\left(\frac{\nu_{\rm syn}}{{\rm GHz}}\right)^{-1/2} ,  \nonumber
\end{align}
where $U_{\rm rad}$ is the energy density in radiation or photons.

For IC losses to be significant, $t_{\rm IC}$ has to be comparable to
$t_{\rm D}$. This condition gives the required $U_{\rm rad}$ as
\begin{align}
  U_{\rm rad} \simeq & 3.33 \times 10^{-9}{\rm\; erg \; cm^{-3} \;}
\left(\frac{R}{3 {\rm kpc}}\right)^{-8/3} \nonumber \\
& \times \left(\frac{l_0}{0.3 {\rm kpc}}\right)^{4/3}
\left(\frac{\nu_{\rm syn}}{{\rm GHz}}\right)^{-2/3} .
\label{urad}
\end{align}
The energy density of the CMB photons is given by
\begin{equation}
 U_{\rm CMB} \equiv aT^4 \simeq 4 \times 10^{-13} (1+z)^4 {\rm \;erg\;cm^{-3}}.
\end{equation}
This gives the time scale for IC losses against CMB photons as
\begin{align}
 t_{\rm CMB} = & \frac{E}{|dE/dt|_{\rm IC}} \\
\simeq & 2.55\times10^8 {\rm\;yr\;}
\left(1+z\right)^{-3}
\left(\frac{\nu_{\rm syn}}{{\rm GHz}}\right)^{-1/2} . \nonumber
\end{align}
Comparing this to $t_{\rm D}$, we get the redshift, at which the electrons
lose a significant fraction of their energy to CMB photons, as 
\begin{align}
  1+z \simeq 9.54 & \times
\left(\frac{R}{3 {\rm kpc}}\right)^{-2/3}
\left(\frac{l_0}{0.3 {\rm kpc}}\right)^{1/3} \nonumber \\
& \times \left(\frac{\nu_{\rm syn}}{{\rm GHz}}\right)^{-1/6} .
\end{align}
This may be a significant effect for very high redshift ($z\sim8.5$)
galaxies. Thus CMB IC cooling of relativistic electrons may have
a role in explaining the systematic lowering of LBG radio fluxes.
However it cannot have any significant role in the case of Peas
which are at $z\sim0.2$.

\citet{1991ApJ...378...65C} suggested that radio emission from
compact starbursts in ultra-luminous infrared galaxies may be suppressed
due to IC losses against the radiation energy density produced by its own
young stellar population. Comparing the IC loss timescale to the electron
diffusion timescale, we have obtained the required energy density in
Eq. \ref{urad}. 
The typical value of $U_{\rm rad}$, required for IC losses to
significantly suppress the synchrotron flux at GHz frequencies, is
much larger than the Milky Way value \citep{1992ARA&A..30..575C} of
$10^{-12} {\rm erg \; cm^{-3}}$ but less than the higher values
of upto $10^{-8} {\rm erg \; cm^{-3}}$ which are seen by
\citet{1991ApJ...378...65C} in compact starbursts.
So it is possible that some
Peas may have suppressed radio fluxes as a result of IC
losses against their own radiation density leading to a wide range
of observed values for $f_{\rm nth}$.

One out of three Peas observed by
GMRT showed no detectable flux (Table \ref{GMRT}).
We therefore estimate
$U_{\rm rad}$ in this galaxy from its SDSS spectra, applying appropriate
bolometric correction for a starburst galaxy. We find an
$U_{\rm rad}\approx4\times10^{-12} {\rm erg \; cm^{-3}}$ which
rules out IC cooling of electrons as the cause for suppression of
radio luminosity. Hence the upper limit of flux from this object
allows us to limit the magnetic field using the diffusion
and equipartition arguments.

\section{Magnetic Field: Diffusion Argument}
\label{magdiff}
{To explain the observed radio emission
from Peas as synchrotron loss from electrons, the magnetic
field must be large so that electrons lose enough energy before
they diffuse out. This can be used to constrain magnetic fields
in Peas.}
Since CMB IC cannot account for the systematic lowering of Pea
radio fluxes we explore
another explanation suggested by \citet{2008ApJ...689..883C} for the
suppression of LBG fluxes. Cosmic
Rays may diffuse easily from systematically smaller galaxies in the early
Universe before they can lose their energy via synchrotron emission.
Since Peas are of sizes comparable to LBGs \citep{2009MNRAS.399.1191C}
this effect should be observable in Peas. Here we use
the observed radio emission to derive the magnetic
field of a Pea, by comparing the electron
diffusion \citep{1964ocr..book.....G} and synchrotron
energy loss timescales \citep{1979rpa..book.....R}.
Thus we get a magnetic field required to explain the
observed radio flux;
\begin{align}
  B_{\rm D} \simeq & 39 {\rm\;\mu G\;}
\left(\frac{R}{3 {\rm kpc}}\right)^{-4/3}
\left(\frac{l_0}{0.3 {\rm kpc}}\right)^{2/3} \nonumber \\
& \times \left(\frac{\nu_{\rm syn}}{{\rm GHz}}\right)^{-1/3}
\left(\frac{f_{\rm nth}}{0.5}\right)^{2/3},
\end{align}
where the magnetic coherence length $l_0$ has a typical
value of 0.3 kpc \citep{2010PhRvL.105i1101C}.
Using $f_{\rm nth}\simeq0.53$ at $\nu_{\rm syn}=1.4$ GHz (from stacking
experiment) and characteristic size \citep{2009MNRAS.399.1191C} $R$ of 3 kpc,
we obtain a typical magnetic field of $B\approx36 {\rm \mu G}$.
Thus the electron diffusion argument provides us with a direct
estimate of the characteristic magnetic field in Peas.

\section{Magnetic Field: Equipartition Argument}
{We show here that the previous estimate of
magnetic field does not lead to absurd energy requirements.}
An independent estimate of magnetic field
can be obtained from energy minimization considerations.
The so called \textit{equipartition magnetic field} can
be estimated by minimizing total energy invoked in relativistic particles
and magnetic fields, in order to explain the observed flux.
The magnetic field corresponding to the minimum energy condition
of \citet{1956ApJ...124..416B}
in a synchrotron plasma can be derived from radio observations.
We follow the derivation of \citet{1993MNRAS.261..445F} with
the typical value of the cosmic ray proton/electron energy
ratio \citep{1964ocr..book.....G} $k=100$, and a
typical spectral index of $\alpha=3/4$ \citep{1993MNRAS.261..445F} for
the radio spectrum, to derive the equipartition magnetic field as
\begin{equation}
   B_{\rm eq} \simeq 49 {\rm\;\mu G\;}
\left(\frac{S_\nu}{{\rm mJy}}\right)^{2/7}
\left(\frac{\theta}{1^{\prime\prime}}\right)^{-4/7}
\left(\frac{\nu_{\rm syn}}{{\rm GHz}}\right)^{3/14} .
\end{equation}
If the actual magnetic field is different from this value, the energy
required to explain the synchrotron emission goes up very rapidly.
It has been argued by \citet{1990IAUS..140..235D} that the real
magnetic field is limited to differ at most by an order of magnitude
above and below the equipartition value derived in this manner.
From the stacking experiment the mean flux of the Peas
is $S_\nu\sim0.12$ mJy.
While they are unresolved at VLA FIRST resolution, serendipitous
HST observations \citep{2009MNRAS.399.1191C} provide their characteristic
size as 3 kpc (where $1^{\prime\prime}$ is $\sim5$ kpc at their
typical redshift).

This fixes the characteristic equipartition magnetic
field of Peas at $B\approx39 {\rm \mu G}$.
The surprising agreement between the estimates of Pea magnetic
fields derived from two independent methods,
points to a characteristic magnetic
field of $\gtrsim30 {\rm \mu G}$ which is
comparable to or even larger than the average Milky Way
value of $B\approx5 {\rm \mu G}$ \citep{1992ARA&A..30..575C}.

{
\section{Discussion}
We report the first radio detection of Peas and show
that this discovery implies large magnetic
fields in Peas, using reasonable assumptions about Cosmic Ray diffusion
and total energy considerations.
}
Given that the Peas are very young galaxies (around one hundredth
the $\approx10^{10}$ yr age of the Milky Way)
{the discovery of radio emission and the implied
$\mu$G magnetic fields}
is a striking result.
Present day magnetic fields are thought to be the
result of amplification of seed fields ($\sim10^{-20}-10^{-18} G$)
by dynamo action \citep{2002RvMP...74..775W} over the galaxy's
lifetime. These models for
the growth of magnetic fields have e-folding times \citep{2002RvMP...74..775W}
of $\approx10^8 - 10^9$ yr. They produce 20 \citep{2008RPPh...71d6901K} to
50 \citep{2006AN....327..395R} e-folds of an efficient dynamo during the
lifetime of the Milky Way. Hence, they can only play a
small role in amplifying the pregalactic \citep{2008RPPh...71d6901K} magnetic
fields within the age of a Pea.
It has been suggested by \citet{1990ApJ...364..451D} that galaxy-wide
$\mu G$ level magnetic fields may be generated by magnetized plasma
produced in jets from a central compact object. Even this
is unlikely as we are considering only the
Star Forming Peas and not the Seyfert Peas. Our results strongly favor the
suggestion from \citet{2008RPPh...71d6901K} that seed fields were
amplified significantly by turbulence (up to $\mu G$ level pregalactic fields)
as protogalaxies and similar substructures formed.

As Peas are close analogs to LBGs it seems likely that young primeval
galaxies could have had strong enough magnetic fields to seed
the intergalactic magnetic field \citep{1999ApJ...511...56K}.
It has been suggested by \citet{1983flma....3.....Z} that magnetic fields
may play a crucial role in star formation. Observed magnetic
fields in nearby galaxies scale slowly with the
SFR \citep{1994ApJ...433..778V} but it is likely that
magnetic fields strongly affect the star formation
efficiency \citep{1999ApJ...517L..69T}. In such a scenario, presence
of strong pregalactic magnetic fields may be the reason why Peas
have started to produce stars at such a large rate. 
Most theoretical and computational studies of reionization era
assume lack of dynamically significant magnetic
fields \citep{2001ARA&A..39...19L}, during formation of first stars.
Our results may require a rethinking of this assumption, in presence
of $\mu G$ level pregalactic magnetic fields. A detailed
study of processes at work in Peas in the nearby
Universe, will help us understand magnetic field
amplification in early galaxies, its role in star formation,
supernova feedback and cosmic ray acceleration.

\acknowledgments

GMRT is run by the National Centre for Radio Astrophysics
of the Tata Institute of Fundamental Research. The VLA is run by the
National Radio Astronomy Observatory, which is a facility of the
National Science Foundation operated under cooperative agreement by
Associated Universities, Inc. We thank Avi Loeb for discussions
and an anonymous referee for comments.










\end{document}